\documentclass[preprint,showpacs,preprintnumbers,amsmath,amssymb,superscriptaddress]{revtex4}
\include{ams}
\usepackage{amsmath,amssymb,amsthm}
\usepackage{dcolumn}
\usepackage{bm}
\usepackage{graphicx}
\usepackage{float}
\usepackage{subfig}

\begin{document}

\title{Multi-agent based analysis of financial data}

\author{T.~Tok\'ar}\email{tomastokar@gmail.com}
\author{D.~Horv\'ath}\email{horvath.denis@gmail.com} 
\author{M.~Hnatich}\email{hnatic@saske.sk} 
\affiliation{SORS Research a.s, 040 01 Kosice, Slovak Republic}

\pacs{89.65.Gh, 89.65.Gh, 05.10.-a, 05.65.+b}

\date{\today}

\begin{abstract}

In this work the system of agents is applied to establish a model of 
the nonlinear distributed signal processing.  The evolution of the system 
of the agents - by the prediction time scale 
diversified trend followers, has been studied for the 
stochastic time-varying environments represented by the real currency-exchange 
time series.  The time varying population 
and its statistical characteristics have been analyzed 
in the non-interacting and interacting cases. 
The outputs of our analysis 
are presented in the form of the mean 
life-times, mean utilities and corresponding 
distributions. They show that populations are susceptible  
to the strength and form of inter-agent interaction.
We believe that our results will be useful for 
the development of the robust adaptive prediction systems.

\end{abstract}

\maketitle 

\section{Introduction}

The agent-based models are distributed computational models which claim 
to treat complex problems by turning them into a multitude 
of the subtasks which can be efficiently solved by means 
of the elementary specialized autonomous entities called agents. 
There are well known examples of the agent-based and multi-agent models
which claim to simulate complex economic~\cite{Palmer94,Shimokawa2007} and
social~\cite{Multi2009} phenomena which are beyond 
traditional analysis in economics. 
Here we discuss entirely different approach and problem. 
We apply agent-based modeling to investigate time series data~\cite{Cao2010}
that have an economic origin.  In such case 
a final answer relevant 
for the investigator of the population may be extracted 
in a way that pieces 
of the output information belonging 
to individual agents are synthesized 
to form the statistical output of interest.  

\subsection{Model of non-interacting entities, population of scale-dependent trend-followers}

    Trend following ({\bf TF}) is an investment strategy that aims to analyze
    and benefit from market trend mechanism. {\bf TF} strategy 
    which focuses on the price moves both up and down is based on assumption 
    of time continuation of the moves. Many variations of this basic framework exist, but here we focus 
    on the consequences of the minimalist, less CPU time demanding approach. 
    Our multi-agent system is built to describe the scale-selection process.  
    We assumed the constant population of $N_{\rm TF}$ agents, 
    where $i$th agent is characterized by its individual operating 
    time scale $l_i \in \langle  l_{\rm min}, l_{\rm max} \rangle$  
    ($i=1, \ldots,  N_{\rm TF}$), 
    which specifies the time window relevant for decisions. 
    As new agent is born, its $l_i$ is chosen randomly within given bounds
    but it does not vary during the agent's lifespan.
    The elementary {\bf TF} agents we deal with  
    are perceiving trend by comparing actual 
    price $p(t)$ with preceding time-shifted $p(t - l_i)$, 
    where $t$ denotes index of tick quotation. The actual agent's sell-buy 
    trade decision is encoded 
    by the variable $s_i(t)$ as it follows:
    \begin{eqnarray}
    s_i(t) &=&
    \left\{ 
    \begin{array}{rcl} 
    +1 \,\,  & \mbox{for \, buy\,  order}   &   p(t) > p(t - l_i),
    \\
    -1  \,\, & \mbox{for \, sell\,  order}  &   p(t) \leq p(t - l_i)
    \end{array} 
    \right.\,. 
    \label{eq:trend1}
    \end{eqnarray}
    In the elementary variant of the model the decisions 
    of {\bf TF} agents are made independently. 
    We analyzed tick-by-tick sequences 
    of the currency exchange rate quotations. By each simulation, population 
    of agents proceeds over the middle price records
    $p(t)$ from chosen data file. 
    We treated $p(t)$ obtained by the standard 
    transformation $p(t) = (p_{\rm ask} + p_{\rm bid})/2$ 
    from the ask and bid tick by tick quotations. 
    The information from $p(t)$ environment 
    is discretized by the following scheme    
    \begin{eqnarray}
    \delta p (t + h) &=&
    \left\{ 
    \begin{array}{lll} 
    +1   & , & p(t)  \leq  p(t + h)\,,
    \\
    -1   & , & p(t)   >   p(t + h)
    \end{array} 
    \right.\,\,. 
    \end{eqnarray}
    At each iteration, the deviation of prediction 
    made for $h$ steps forward 
    is measured by the recurrently 
    given utility $u_i(t + h) =  u_i(t) + s_i(t)\,\delta p(t + h)$. 
    To construct the series of $u_i(t+h)$, 
    it remains to specify initial 
    utility $u_i(t_{\rm initial}) \equiv  u_{\rm born}>0$. 
    The life of the agent begins at the moment of the death 
    of its predecessor labeled by the same index~(i). 
    The agent's life continues until the utility 
    drops to zero bound $u_i=0$.

\section{Results}

  \subsection{The scale-formation process of independent TF agents}

\begin{figure}
      \centering
      \subfloat{\includegraphics[width = 0.5\textwidth]{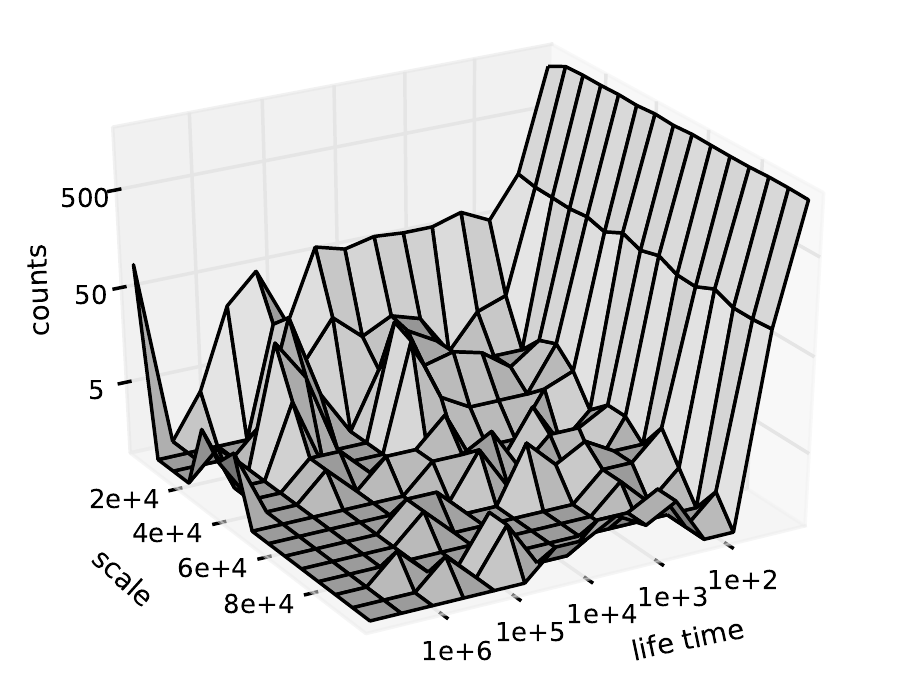}}
      \subfloat{\includegraphics[width = 0.5\textwidth]{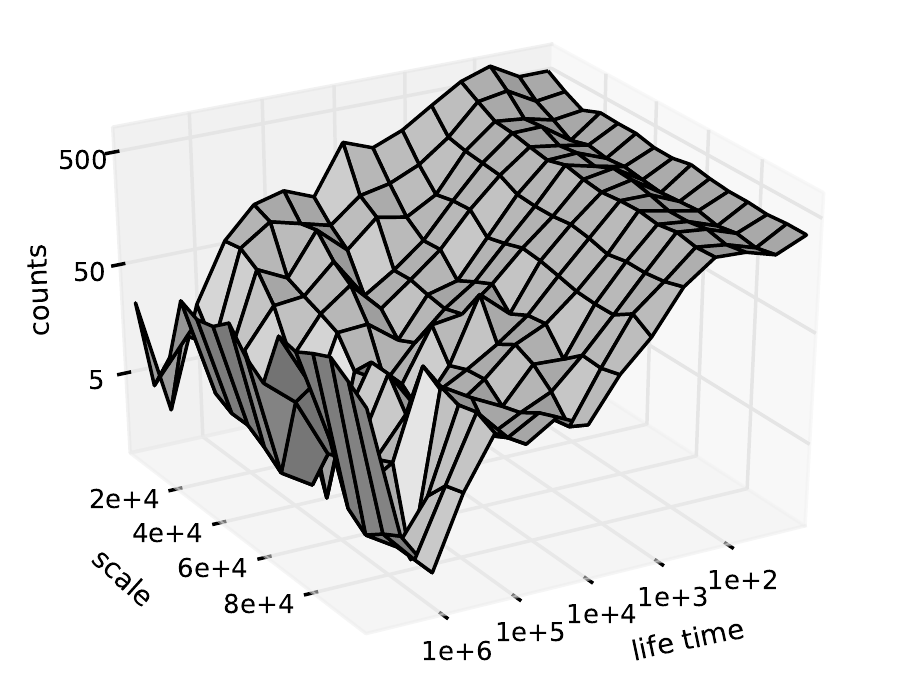}} \\
      \subfloat{\includegraphics[width = 0.5\textwidth]{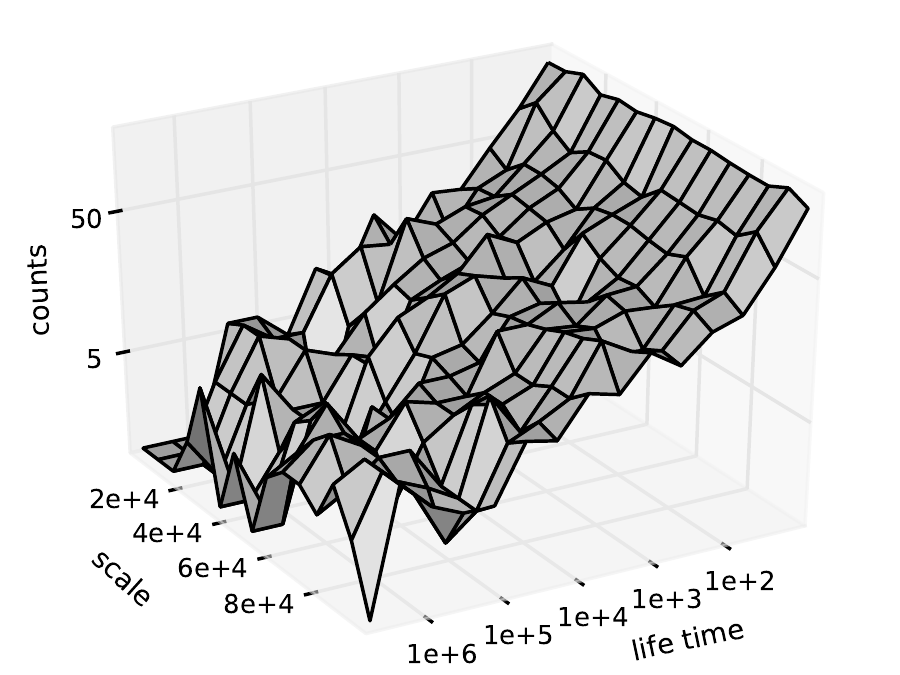}} \\
      \caption{
      Distribution function of the life-times and $l_i$ scales 
      obtained for the 
      population of independent agents. 
      Simulated for EUR/USD 
      exchange rate record.  
      The predictions are made for 
      $h=1,100,1000$ steps forward.}
      \label{3D_EUR_vs_USD1}
      \end{figure}

    We realize that the question of the population size is apparently 
    irrelevant for the independent (non-communication, non-interacting) {\bf TF} agents, 
    however it starts to be important when the aspects 
    of evolution 
    and inter-agent interaction (information exchange)  
    are taken into account.
    The numerical analysis of the population of independent {\bf TF} agents 
    has been done for $N_{\rm TF} = 1000$, $u_{\rm born} = 10$, $h=1$. 
    At the moment of birth, the scale $l_i$ 
    has been 
    drawn randomly 
    uniformly within the 
    bounds $l_{\rm min}= 1$,  
    $l_{\rm max}=10^5$.   
    The approach has been applied to treat the 
    six year record of the EUR/USD 
    exchange rate (period 2004-2009). 
    We have been interested in probability distributions 
    as functions life-time and $l_i$ scales 
    (see Fig.\ref{3D_EUR_vs_USD1}). 
    Apart from the excessive simplicity of the {\bf TF} mechanism, we found 
    that several agents survived after the complete data record had been 
    browsed. Then, what follows from it is that {\bf TF} strategies may provide 
    adequate ground for some basic prediction models. 
    We extended simulation 
    to include USD/CHF, USD/JPY records. 
    Similar, but not identical distributions have 
    been obtained for these data. 
    We can state that {\bf TF} agents provide a better prediction 
    (for $h=1$) if they focus on the small, i.e. $l_i\sim 1$ scales. The statement is general 
    and covers all aforementioned foreign exchange records. 
    Next question we claim to answer is "How the distribution of life-time
    and $l_i$ will be affected, when the prediction is made for more 
    than one step 
    forward ($h > 1$)?". 
    Let's compare previous 
    results for $h=1$ with those obtained for $h = 100$. The results depicted 
    in Fig.\ref{3D_EUR_vs_USD1} appear to imply 
    that no general statement regarding 
    life-time 
    can be made. There is, however, 
    some qualitative difference which might be noticed, that 
    functional relation between life-times 
    and $l_i$ seems to be less readily seen for 
    $h=100$ than it is in the case of $h=1$. 
    As $h = 1000$, the results become  
    more currency specific. 
    One may conclude that in general, 
    the increase in $h$ leads to the inhibition 
    of preferential 
    scales.

  \subsubsection{Populations of agents - instant averages and transient characteristics}

    \begin{figure}
      \centering
      \subfloat{\includegraphics[width = 0.5\textwidth]{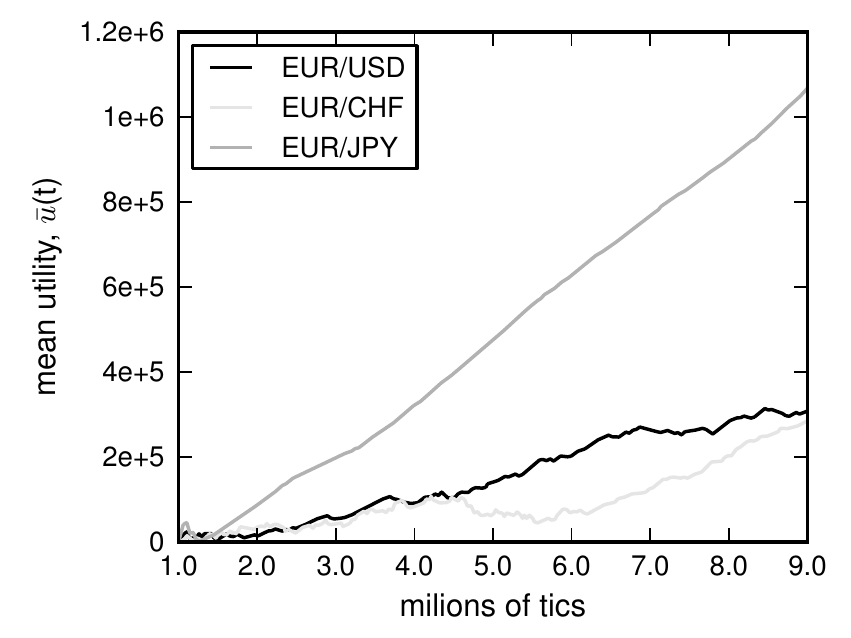}}
      \subfloat{\includegraphics[width = 0.5\textwidth]{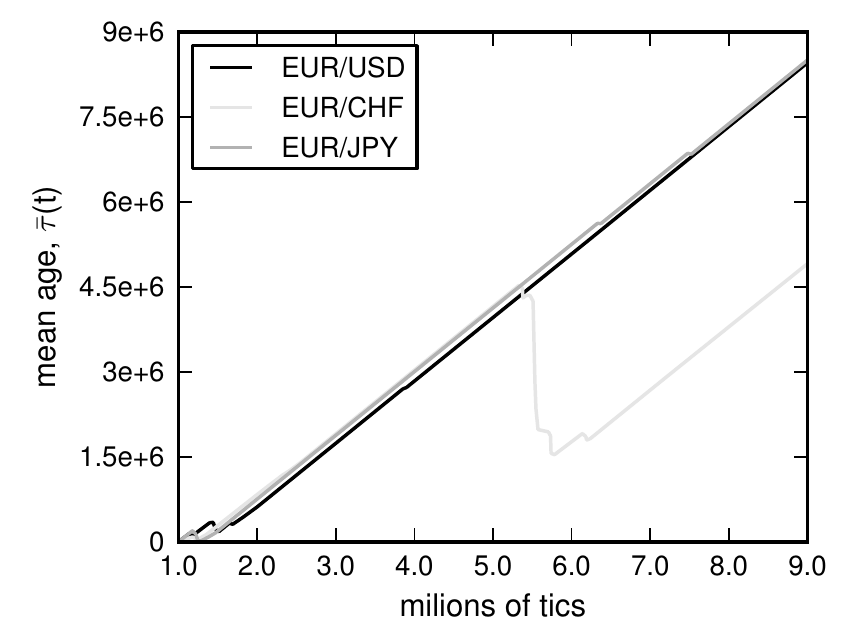}} \\
      \subfloat{\includegraphics[width = 0.5\textwidth]{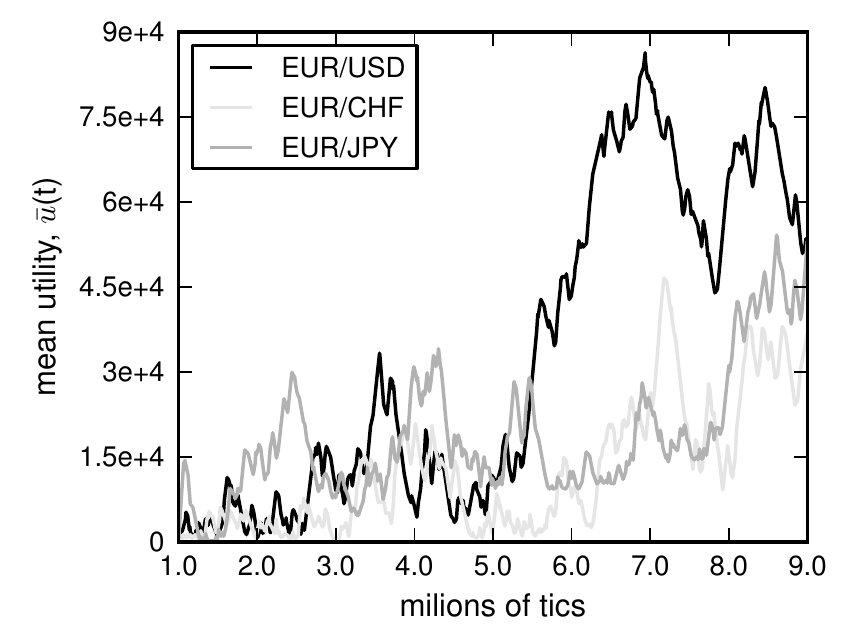}}
      \subfloat{\includegraphics[width = 0.5\textwidth]{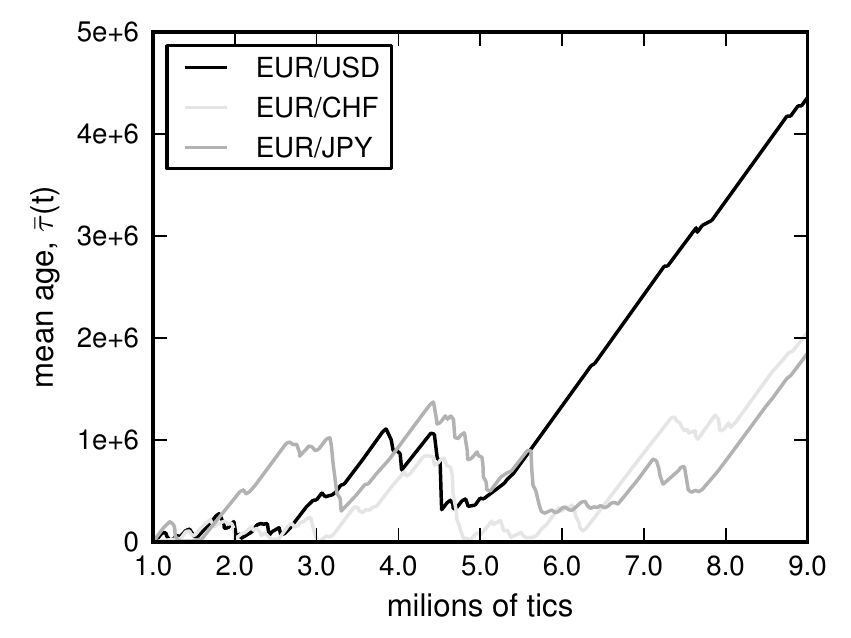}} \\
      \caption{The monitoring of the transient 
      regime by means of 
      $\overline{u}(t)$ and $\overline{\tau}(t)$ instant averages.
      The results 
      obtained for 
      different $h$ 
      and different currencies; 
      $h = 1$ (top), 
      $h= 100$ (bottom). 
      One may see more 
      thorough 
      "mixing" of mean utilities at higher $h=100$.}  
      \label{non-interacting_tranzient}
    \end{figure}
    The agents are predicting, 
    gaining or losing pieces 
    of their utilities, some agents are dying and some new agents 
    are being born. 
    To shed light on the population 
    dynamics we calculated transient characteristics 
    including temporal development 
    of two instantaneous population averages:
    mean utility $\overline{u}(t)= \sum_{i=1}^{N_{\rm TF}}/N_{\rm TF}$ 
    and 
    mean actual age 
    $\overline{\tau}(t) = \sum_{i=1}^{N_{\rm TF}} \tau_i(t)/N_{\rm TF}$, 
    where $\tau_i(t)$ represents actual age of 
    agent $i$ at tick $t$.   
    As can be seen from Fig.\ref{non-interacting_tranzient}, 
    the mean utility is growing faster when the predictions 
    are made for smaller number of steps forward ($h\sim 1$). 
    Within the EUR/USD record, 
    after the initial transient $t_1 = 9 \cdot 10^6$ ticks, 
    the accumulated $\overline{u}(t_1)$ reaches values 
    around $3.5 \cdot{10^{5}}$,  $5\cdot{10^{4}}$,  $1\cdot{10^{4}}$, 
    for the predictions $h=1, 100$  
    steps forward, respectively.
    However, if $\overline{u}(t_1)$ is measured for other currencies,
    the comparatively slower growth 
    has been observed as $h$ increases.
    The value $\overline{u}(t_1)$ 
    may be used as well 
    to define the prediction accuracy 
    by means of the formula
    ${\rm PA}=( t_1 +  \overline{u}(t_1))/(2 t_1)$.
    The calculation of ${\rm PA}$ leads
    $52\%$ - $55\%$ (for $h=1$), 
    $50.05\%$ - $50.3\%$ (as $h=100$) and 
    $50.005\%$ - $50.1\%$ (as $h=1000$ ticks). 
    The monitoring 
    of $\overline{u}(t)$ and $\tau(t)$ 
    revealed stages of the continuous growth interrupted 
    by the bursts consisting of several 
    sudden small drops.  
    These events could 
    be interpreted 
    in terms of the population biology as massive 
    {\em extinctions}. The extinctions may be also be perceived 
    as an expulsion of {\bf TF} agents 
    from certain $l_i$ scales 
    and determination of temporary stableness 
    at some another specific scales. 
    The observation of the subsequent renewal 
    of the regime with less 
    fluctuating $\overline{u}(t)$ and 
    $\overline{\tau}(t)$ 
    result in epochs of the temporarily sustainable generations.  

     \subsection{The information exchange between agents}

    Until now we have studied idealized population of elementary agents 
    where each member acts 
    independently of others. It seems interesting to extend our study  
    to strategies which are based on the inter-agent interaction. 
    The experience from the design of the complex systems 
    is that interaction may lead to the emergence of the 
    unexpected phenomena. Our hypothesis is that under 
    certain conditions 
    the interactions of agents may extenuate 
    undesired data effects
    (like those which cause massive extinctions), or may improve 
    the quality of the
    predictions. Within the next subsections we discuss several
    models of interaction which extend picture 
    of the {\bf TF} agent in a very natural way. 
 
    \subsubsection{Superior - merchant agents with exceptional information access.}

      The elementary evolutionary belief we are implementing 
      lies in the fact that  exceptional access to information resources 
      may to enhance the performance of decisions.  
      In this respect, we want to supplement population by better 
      informed decision makers. We have introduced and tested effect of agents 
      we shall call {\bf merchants} or {\bf M} agents. 
      They are designed to receive information about 
      the pure expected {\bf TF} 
      decisions or to provide information to assist 
      {\bf TF} decisions modified by the recommendation process. 
      The information {\bf M}-agent can get from $i-th$ {\bf TF} agent is always 
      reduced to the  instant $(u_i(t), \tau_i(t))$ pair. 
      The comparative process within the population yields decision output 
      $s_{\rm M}(t)\in \{-1,1\}$ of {\bf M} agent.  More concretely, we studied 
      information processing where {\bf M} agent decides in accordance 
      with a decision of {\bf TF} agent attaining highest instant utility within
      {\bf TF} population.   
      Similar strategies, such as weighted averaging according utility, have been tested 
      as well, however, their contribution does not differ in principal 
      from the results obtained for the above defined strategy of {\bf M} agent. 

    \subsubsection{BM-TF strategy: TF agent born by the M agent.}

      Originally, the non-interacting agents were born with scale drawn 
      uniformly randomly. The evolutionary feature we analyzed 
      is that agents are born 
      at the proximity to the scale actually occupied by the {\bf M} agent. 
      The diversity which prevents from 
      getting stuck is achieved by the mutation process 
      where shift 
      with respect to the scale 
      of {\bf M} agent is modelled 
      by the Gaussian distribution, 
      with the mean localized at the scale of 
      {\bf M} agent (the dispersion is set to be equal to 3000). 
       
    \subsubsection{RM-TF strategy: decisions of TF agents regarding recommendations of M agents.}
     
      As mentioned above, the main motivation for the use 
      of multi-agent data analysis 
      represents an information inter-agent exchange which 
      may induce 
      self-organized behavior and emergence of the qualitative
      unexpected changes in the information treatment. 
      More specifically, by means of correlated decisions 
      we claim to induce self-improvement 
      and stability of the {\bf TF} population. 
      With this aim we implemented information exchange 
      where agents 
      are comparing their predictions with 
      recommendations concerned in the unique central variable 
      $s_{\rm M}(t)$. 
      The information in the form of $s_{\rm M}(t)$ is transmitted 
      by the {\bf M} agents which are endowed 
      with an exceptional authority to affect subordinated {\bf TF} agents 
      by the {\em recommendation process}.  
      As a consequence, each {\bf TF} agent justifies its preliminary 
      decision $s_i^{\rm pr}$ (calculated according Eq.(\ref{eq:trend1})
      within the scheme of independent agents) with $s_{\rm M}(t)$. 
      The approval scheme for subordinated 
      {\bf TF} agent yields decision model    
      \begin{eqnarray}
	s_i (t)
	&=&
	\left\{ 
	\begin{array}{lll} 
	s_i^{\rm pr} (t)\,   &  ,  &   s_i^{\rm pr} (t)  =    s_{\rm M}(t)\,,
	\\
	0                    &  ,  &   s_i^{\rm pr} (t) \neq  s_{\rm M}(t)
	\end{array} 
	\right.\,. 
      \end{eqnarray}
      The formula gives rise to the 
      newly defined {\em passive state} $0$ which stems from
      disagreement of agents. The proposed here {\bf RM-TF} strategy closes 
      the information {\em feedback loop} since the {\bf M} agent has 
      access to decisions of subordinated {\bf TF} community. 
      The results of simulation are presented in
      Fig.\ref{comparison_tranzient}. They show the influence
      of the interaction effect on the stability of population.  

  \subsection{The statistical consequences of interaction}

    \begin{figure}
      \centering
      \subfloat{\includegraphics[width = 0.44\textwidth]{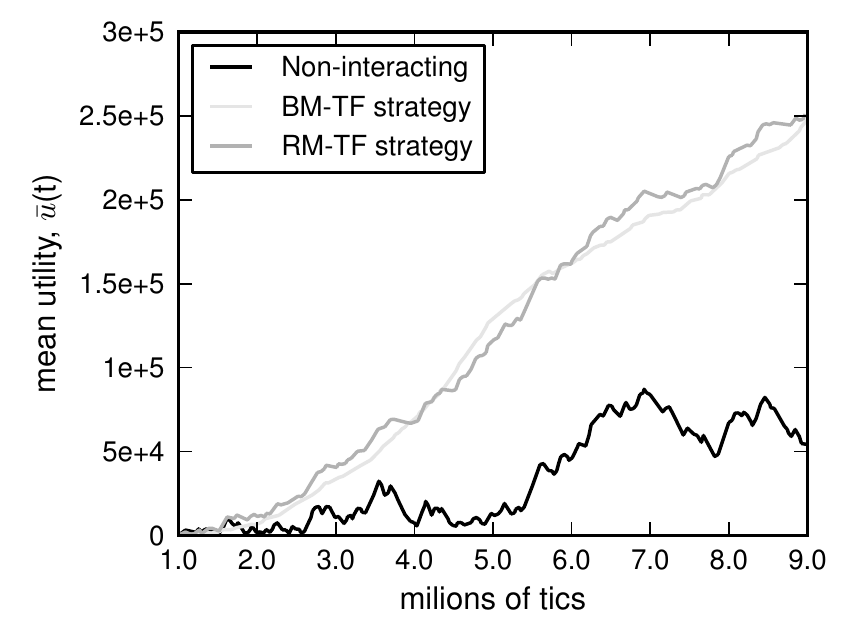}}
      \subfloat{\includegraphics[width = 0.44\textwidth]{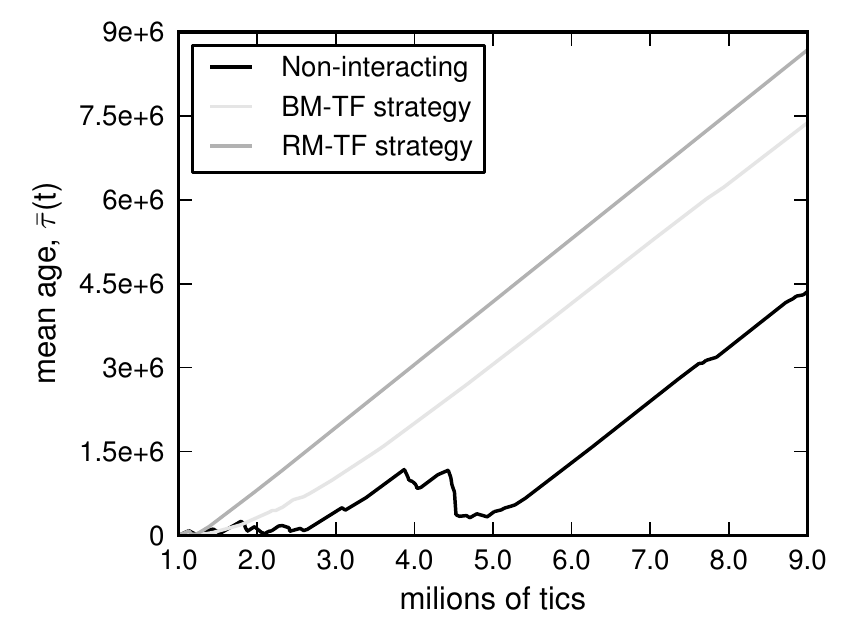}} \\
      \caption{The comparison of the transient characteristics 
      measured for the non-interacting and interacting agents~(for {\bf BM-TF, RM-TF} strategies, $h=100$).}
      \label{comparison_tranzient}
    \end{figure}
    \begin{table}
    \begin{center}
    \caption{The list of the log-log slopes (effective indices) 
    of the distributions from Fig.\ref{comparison_distributions}.} 
    \begin{tabular}{|lll|}
    \hline
    distribution of life-times  & range   &  effective index   \\ 
    \hline
    non-interacting  & $ <100, 10000>$    & -0.50     \\ 
    BM-TF            & $ <17,10000>$      & -0.30     \\
    RM-TF            & $ <100,10000>$     & -0.40     \\
    \hline
    \hline     
    distribution of deathrates   & range  & effective index 
    \\  
    \hline
    non-interacting  & $<1, 11.27>$  &  -0.73 \\
    BM-TF            & $<1, 11.64>$  &  -0.83 \\ 
    RM-TF            & $<1, 11.75>$  &  -0.50 \\
    \hline
    \end{tabular}
    \label{table1}
    \end{center}
    \end{table}
    \begin{figure}
      \centering
      \subfloat{\includegraphics[width = 0.5\textwidth]{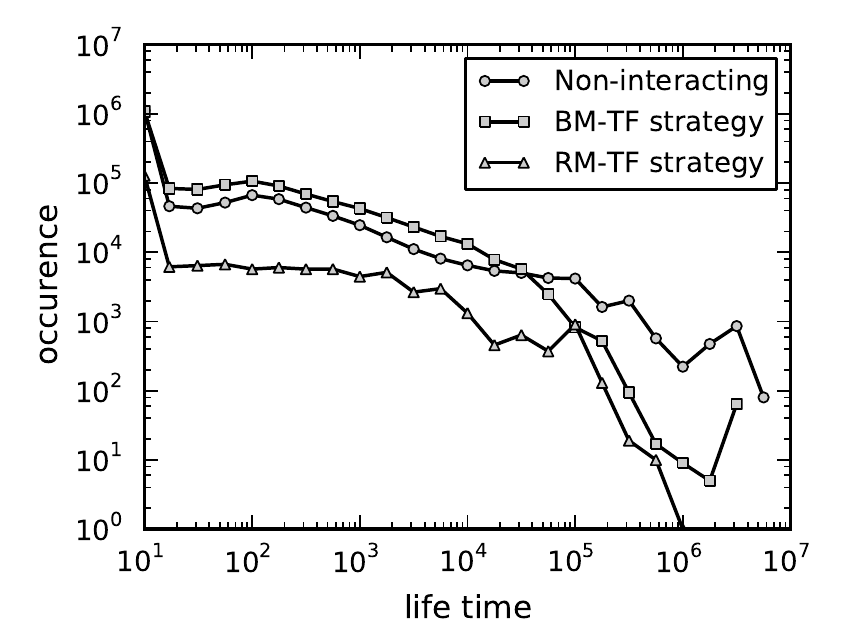}}
      \subfloat{\includegraphics[width = 0.5\textwidth]{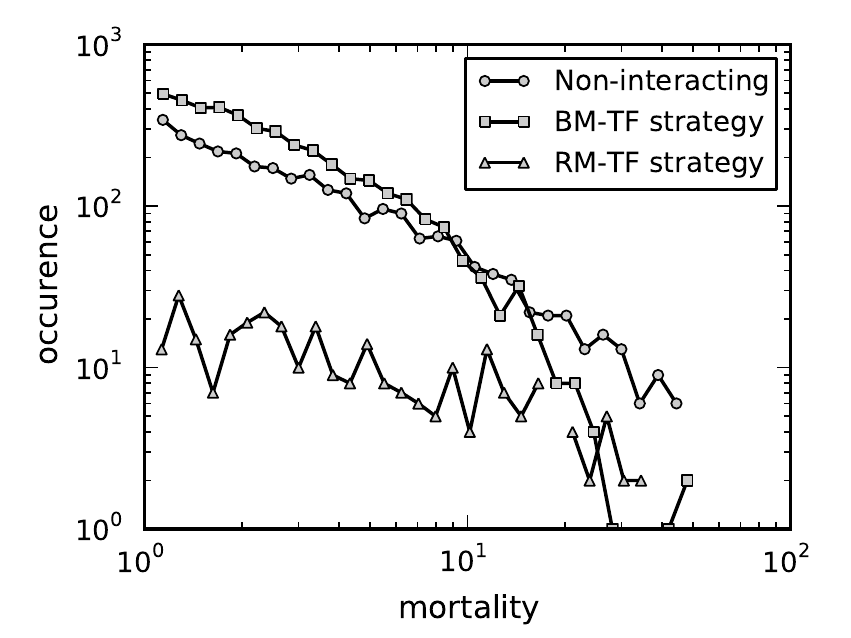}} \\
      \caption{The log-log plot of distributions of the life-times (left) 
      and number of deaths
      (mortalities, deathrates) 
      per tick (right)
      determined for different 
      strategies. 
      Calculated for 
      $h=100$. Most remarkable violation 
      of the power-law form of distributions occurs in the case of {\bf RM-TF} interaction.
      The indices~(exponents) of nearly power-law 
      distributions are listed 
      in the table~\ref{table1}.}
      \label{comparison_distributions}
    \end{figure}

    We compared characteristics of non-interacting population 
    with those obtained by means of the combined {\bf BM-TF}, {\bf RM-TF} 
    strategies (see Fig.\ref{comparison_tranzient}). These findings imply statistical stabilization 
    effect of the interaction. The transient epochs  monitored by means of $\overline{u}$ and $\overline{\tau}(t)$ clearly indicate that massive 
    extinctions of the non-interacting {\bf TF} population may be remarkably reduced upon the implementation 
    of the {\bf BM-TF, RM-TF} types of the information exchange. 
    From the point of view of the universal features of the distributions and their hypothetical power-law form      
    (see Fig.\ref{comparison_distributions}), we observed a remarkable 
    effect of the combined strategies on the distributions of the life-times. 
    The population of {\bf TF} agents gains nearly critical properties~(see
    e.g.~\cite{Honga2007}) indicated by the power-law form with the indices 
    (log-log slopes or exponents) listed in table~\ref{table1}.  The dynamics of given system 
    have been found to be similar to those of complex systems, conventionally 
    studied in physics of critical phenomena. 
    In contrast, we uncover that inter-agent {\bf RM-TF} type interaction ruins the power-law 
    form pertinent to currency data. However, this finding most typical for EUR/USD 
    differs for other currencies.  Of course, artificially suggested forms of interaction   
    seem to be very far from the interaction of the real market agents, 
    therefore, there may appear loose of the power-law form.

  \section{Conclusions} 

  We studied model of data analysis based on the system of autonomous agents. 
  The model has been tested for 
  several selected currencies 
  in the case of the non-interacting (independent) 
  and interacting agents. 
  The modelling of the non-interacting agents 
  uncovered specific periods of the extinctions of agents 
  (and belonging scales) interrupted  
  by the relatively calm periods. 
  It has been demonstrated that specific 
  inter-agent interaction may stabilize population of agents  
  and inhibit in part criticality which is typical 
  for financial market data.

\end{document}